\documentclass[aps,pre,preprint]{revtex4-1} 
\usepackage{amsfonts}
\usepackage{amsmath}
\usepackage{amssymb}
\usepackage{graphicx}
\usepackage{mathrsfs} 
\usepackage{mathrsfs}
\usepackage{color} 
\usepackage{graphicx}
\usepackage{afterpage}

\providecommand\bnabla{\boldsymbol{\nabla}}

\newcommand{\pd}[2]{\frac{\partial #1}{\partial #2}}

\begin{document} 
\title{Singular effective slip length for longitudinal flow over a dense bubble mattress}
\author{Ory Schnitzer}
\affiliation{Department of Mathematics, Imperial College London, South Kensington Campus, London, United Kingdom}

\begin{abstract}
We consider the effective hydrophobicity of a periodically grooved surface immersed in liquid, with trapped shear-free  bubbles protruding between the no-slip ridges at a $\pi/2$ contact angle. Specifically, we carry out a singular-perturbation analysis in the limit $\epsilon\ll1$ where the bubbles are closely spaced, finding the effective slip length (normalised by the bubble radius) for longitudinal flow along the the ridges as 
${\pi}/{\sqrt{2\epsilon}}-(12/\pi)\ln 2 + (13\pi/24)\sqrt{2\epsilon}+ o(\sqrt{\epsilon})$, the small parameter $\epsilon$ being the planform solid fraction. 
The square-root divergence highlights the strong hydrophobic character of this configuration; this leading singular term (along with the third term) follows from a local lubrication-like analysis of the gap regions between the bubbles, together with general matching considerations and a global conservation relation. The $O(1)$ constant term is found by matching with a leading-order solution in the ``outer'' region, where the bubbles appear to be touching. We find excellent agreement between our slip-length formula and a numerical scheme recently derived using a ``unified-transform'' method (D. Crowdy, \textit{IMA J. Appl. Math.}, \textbf{80} 1902, 2015). The comparison demonstrates that our asymptotic formula, together with the diametric  ``dilute-limit'' approximation (D. Crowdy, \textit{J. Fluid Mech.}, \textbf{791 R7}, 2016), provides an elementary analytical description for essentially arbitrary no-slip fractions. 
\end{abstract}

\maketitle
\section{Introduction}
There is great current interest in the design and application of micro-structured ``meta-surfaces'' that are effectively superhydrophobic \cite{Cottin:03,Choi:06,Truesdell:06,Hyvaluoma:08,Rothstein:10}; flows varying on scales large compared with the microstructure appear to slip over the surface, rather than satisfy a no-slip condition. A wide body of theoretical literature now exists covering general properties \cite{Ybert:07,Kamrin:11,Schmieschek:12}, along with computations and analytic results for the effective slip length of specific micro-structured geometries and materials  \cite{Lauga:03,Davis:09,Teo:10,Kamrin:10,Schonecker:13,Enright:14}. Building on the pioneering solutions of Phillip \cite{Philip:72}, a plethora of new results have recently been obtained using complex-variable techniques, in particular conformal mappings \cite{Crowdy:10,Crowdy:11b} and the ``unified-transform'' method \cite{Crowdy:15,Crowdy:15b}. The available numerical and analytical solutions have been further extended by regular-perturbation schemes for nearly flat meniscuses, nearly shear-free inclusions, and well-separated micro-stuctured elements \cite{Sbragaglia:07,Crowdy:16}.

\begin{figure}[b]
\begin{center}
\includegraphics[scale=0.75]{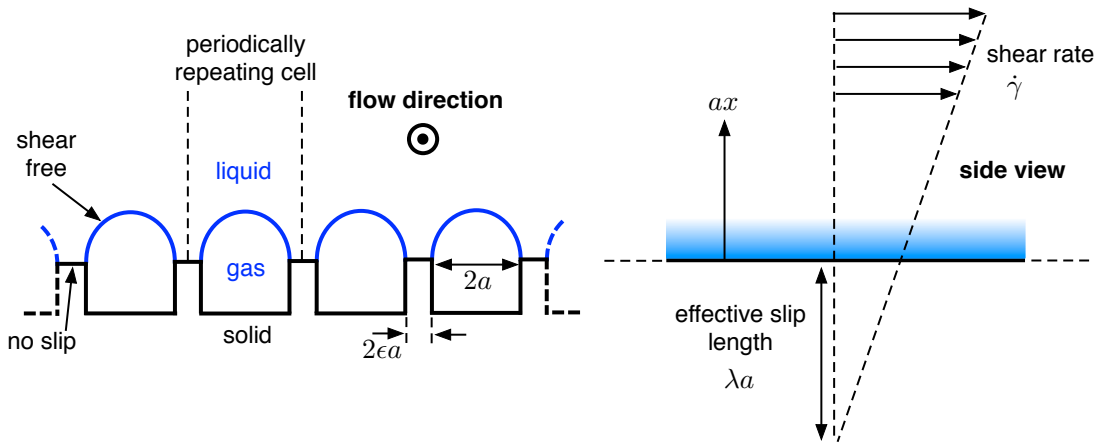}
\caption{Schematic of the slip-length problem for longitudinal flow over a ``bubble mattress''.}
\label{fig:dim_schem}
\end{center}
\end{figure}
A prevalent realisation of a superhydrophobic surface consists of a periodically grooved solid surface immersed in water, with trapped-air pockets protruding between the solid ridges. For this configuration, sometimes termed a ``bubble mattress'' \cite{Crowdy:10}, the effective slip length diverges with vanishing solid fraction $\epsilon$ (at least as long as the air bubbles remain stably trapped). According to the scalings suggested by Ybert \textit{et al.} \cite{Ybert:07}, this divergence is logarithmic, i.e.~for $\epsilon\ll1$ the slip length is commensurate to the product of the periodicity and $\ln(1/\epsilon)$; for macroscopic flows varying on a scale much larger than the surface periodicity, this implies an inherently weak hydrophobic effect. Fortunately, numerical computations hint that the logarithmic scaling breaks down when the meniscuses of the protruding bubbles are appreciably non flat. In particular, for longitudinal flow along the cylindrical bubbles, plots of the slip length against bubble separation depict a rapid growth with vanishing separation \cite{Teo:10,Ng:11}. This is most pronounced in the case of a $\pi/2$ contact angle, see e.g.~Fig.~12 in \cite{Crowdy:15b}. In this paper we carry out an asymptotic analysis of the small-solid-fraction limit $\epsilon\to0$ for $\pi/2$ contact angles. Our goal is to derive an accurate  asymptotic expansion for the effective slip length and thereby highlight the surprisingly large slip lengths attainable with densely grooved surfaces.
%


\section{Problem formulation}
A schematic of the problem is shown in Fig.~\ref{fig:dim_schem}. A periodic array of cylindrical shear-free bubble protrusions (radius $a$ and contact angle $\pi/2$), separated by flat no-slip solid boundaries of thickness $2\epsilon a$, is exposed to a shear flow (shear rate $\dot\gamma$) parallel to the cylindrical bubbles; we assume small capillary numbers and accordingly approximate the bubble cross-sectional boundaries by semicircles.  For unidirectional flow parallel to the applied shear, and in the absence of a pressure gradient, the flow velocity satisfies Laplace's equation, and at large distances is $\sim \dot\gamma a (x + \lambda )$, $ax$ being the normal distance from the solid segments and $a\lambda$ the effective slip length \cite{Crowdy:10}. The problem is periodic and it is sufficient to consider a single ``unit cell'' of width $2a(1+\epsilon)$. 

\begin{figure}[t]
\begin{center}
\includegraphics[scale=0.45]{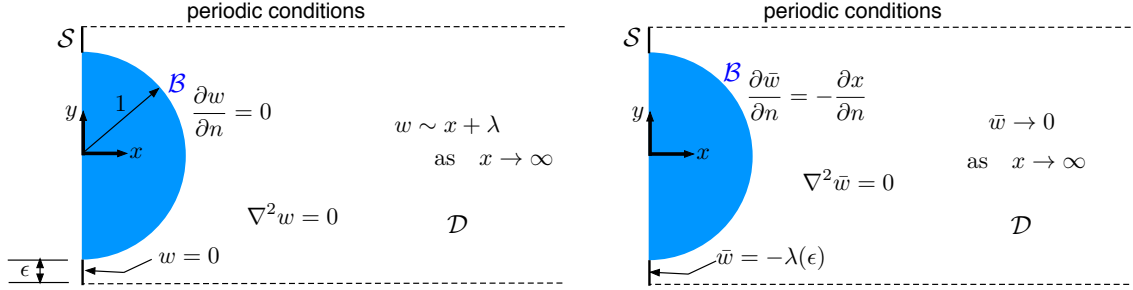}
\caption{Left panel: Dimensionless formulation. The normalised slip length $\lambda(\epsilon)$ is an outcome of the solution to the boundary value problem. Right panel: Formulation in terms of the disturbance velocity $\bar{w}=w-x-\lambda$.}
\label{fig:nondim}
\end{center}
\end{figure}

We adopt a dimensionless formulation where lengths are normalised by $a$ and velocities by $\dot\gamma a$, and define a Cartesian co-ordinate system $(x,y)$, where $y$ is measured from the centre of an arbitrarily chosen bubble. The unit-cell domain $\mathcal{D}$ is thus bounded by $y=\pm(1+\epsilon)$, the bubble interface $\mathcal{B}$, and the flat solid boundaries $\mathcal{S}$. The problem governing the longitudinal velocity component $w$  is depicted in Fig.~\ref{fig:nondim}, and consists of Laplace's equation,
\begin{equation}\label{lap 1}
\nabla^2 w =0 \quad \text{in} \quad \mathcal{D};
\end{equation}
the no-shear condition 
\begin{equation}\label{bubble bc}
\pd{w}{n} = 0 \quad \text{on} \quad \mathcal{B};
\end{equation}
the no-slip condition 
\begin{equation}\label{solid bc}
w=0 \quad \text{on} \quad \mathcal{S};
\end{equation}
the far field condition
\begin{equation}\label{far}
w \sim x + \lambda + o(1) \quad \text{as} \quad x\to \infty;
\end{equation}
and periodic boundary conditions at $y=\pm (1+\epsilon)$. Since $w(y)=w(-y)$ in $\mathcal{D}$ the latter can be equivalently replaced by the Neumann conditions
\begin{equation}\label{period}
\pd{w}{y}=0 \quad \text{at} \quad y=\pm (1+ \epsilon).
\end{equation}
It will prove useful to also keep in mind the integral relation
\begin{equation}\label{int w}
\int_{1}^{1+\epsilon}\left.\pd{{w}}{x}\right|_{x=0}\,dy= 1+\epsilon,
\end{equation}
which is readily derived by integrating Laplace's equation \eqref{lap 1} over $\mathcal{D}$ and applying the divergence theorem. Physically, \eqref{int w} represents the fact that in the absence of a longitudinal pressure gradient or body force the shear force away from the surface is the same as that acting on its solid segments. In what follows, it is helpful to alternatively interpret \eqref{int w} as an integral conservation law with respect to the fictitious irrotational ``flow'' $\bnabla w$. 

In \eqref{far}, the first term corresponds to the prescribed shear, whereas $\lambda(\epsilon)$ is unknown. Our goal is thus to determine $\lambda(\epsilon)$ in the limit $\epsilon\to0$. 
First, however, we reformulate the problem in terms of the disturbance velocity $\bar{w}=w-x-\lambda$, which turns out to be convenient for the asymptotic analysis. The new problem, also depicted in Fig.~\ref{fig:nondim}, is similar to that governing $w$, but with 
condition \eqref{bubble bc} replaced by 
\begin{equation}\label{bubble bar}
\pd{\bar w}{n} = -\pd{x}{n} \quad \text{on} \quad \mathcal{B}; 
\end{equation}
condition \eqref{solid bc} by
\begin{equation}\label{solid bar}
\bar w=-\lambda(\epsilon) \quad \text{on} \quad \mathcal{S};
\end{equation}
and \eqref{far} by 
\begin{equation}\label{far bar}
\bar w \to 0 \quad \text{as} \quad x\to \infty.
\end{equation}
Finally, in terms of $\bar{w}$, the integral relation \eqref{int w} becomes
\begin{equation}\label{intB}
\int_{1}^{1+\epsilon}\left.\pd{\bar{w}}{x}\right|_{x=0}\,dy= 1.
\end{equation}

\section{Closely spaced bubbles}\label{sec:asym}
\subsection{Singular scaling of the effective slip length}\label{ssec:scaling}
Henceforth we consider the asymptotic limit where $\epsilon\to0$. We expect the normalised slip length $\lambda$ to diverge in this limit, but at what rate? The integral relation \eqref{intB} shows that, for arbitrarily small $\epsilon$, there is a finite $O(1)$ ``flux'' $\bnabla\bar w$ through the solid boundaries
 $\mathcal{S}$. Noting that the width of those boundaries is $O(\epsilon)$, and adjacent to them $\bar{w}=O(\lambda)$ [cf.~\eqref{solid bar}], this implies that $\epsilon \lambda/\delta  =O(1)$, where $\delta$ is the length scale on which $\bar{w}$ varies in the $x$ direction close to $\mathcal{S}$. The latter subdomain of $\mathcal{D}$ is  geometrically narrow; in particular, owing to the locally parabolic boundary shape, the separation between the bubbles remains $O(\epsilon)$ for $x=O(\epsilon^{1/2})$. This implies that $\bar{w}$ is approximately independent of $y$ there, and that the right-hand side of \eqref{bubble bar} is small; thus the product of $d\bar{w}/dx$ and the gap thickness is conserved [cf.~\eqref{gap W12}]. But the locally parabolic geometry means that the relative thickness variation is $O(1)$ over a length scale $\epsilon^{1/2}$, i.e.~$\delta =O(\epsilon^{1/2})$. It follows that $\lambda$, and hence $\bar{w}$ in the region between the nearly touching bubbles, both scale like $\epsilon^{-1/2}$. 

\subsection{``Inner'' gap and ``outer'' bubble-scale expansions}\label{ssec:inner_outer}
\begin{figure}[t]
\begin{center}
\includegraphics[scale=0.5]{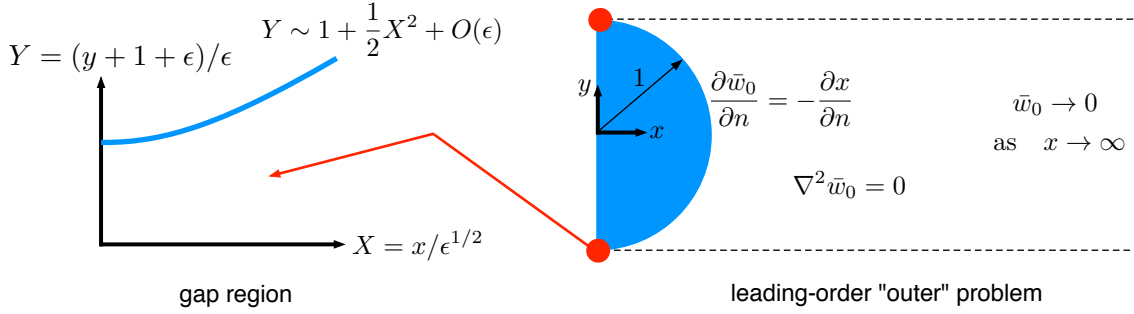}
\caption{Left panel: Stretched co-ordinates used to analyse the gap region.  Right panel: Leading-order outer problem. Matching with the gap regions is required in the limits where $(x,y)\to(0,\pm1)$ from within the outer liquid domain.}
\label{fig:inner_outer}
\end{center}
\end{figure}

The above discussion implies that the asymptotics of $\bar{w}$ as $\epsilon\to0$ are spatially nonuniform. Accordingly, we conceptually decompose the liquid domain into two ``inner'' gap regions, at distances $O(\epsilon^{1/2})$ from the $O(\epsilon)$-thick solid boundaries, and an ``outer'' region away from the gaps, where to leading order the bubbles appear to be touching (see Fig.~\ref{fig:inner_outer}). In preparation for our analysis of the inner region (say, in $y<0$), we define the stretched gap coordinates 
\begin{equation}\label{gap}
Y=(y+1+\epsilon)/\epsilon, \quad X=x/\epsilon^{1/2},
\end{equation}
in which terms the bubble boundary is $Y=H(X) \sim H_0(X)+\epsilon H_1(X) + o(\epsilon)$, where $H_0=1+\frac{1}{2} X^2$ and $H_1=X^4/8$, and a gap disturbance velocity $\bar W(X,Y)=\bar w(x,y)$. The inner problem governing $\bar{W}$ consists of Laplace's equation
\begin{equation}\label{gap lap}
\epsilon \pd{^2\bar W}{X^2}+\pd{^2\bar{W}}{Y^2} = 0, \quad \text{for} \quad 0<Y<H(X), \quad X>0,
\end{equation}
together with the conditions
\begin{equation}\label{gap solid}
\bar{W} = -\lambda(\epsilon) \quad \text{at} \quad X=0,
\end{equation}
\begin{equation}\label{gap periodic}
\pd{\bar{W}}{Y} = 0 \quad \text{at} \quad Y=0,
\end{equation}
and 
\begin{equation}\label{gap bubble}
\pd{\bar{W}}{Y}-\epsilon \frac{dH}{dX}\pd{\bar{W}}{X} - \epsilon^{3/2}\frac{dH}{dX}= 0 \quad \text{at} \quad Y=H(X).
\end{equation}
In addition $\bar{W}$ must match with the outer region as $X\to\infty$. Recall that we also have at our disposal the global relation \eqref{intB}, which now reads as 
\begin{equation}\label{int gap}
\epsilon^{1/2}\int_{0}^{1}\left.\pd{\bar{W}}{X}\right|_{X=0}\,dY= 1.
\end{equation}

In agreement with the scaling arguments given before, it follows from \eqref{int gap} that $\bar{W}=O(\epsilon^{-1/2})$, suggesting the gap expansion 
\begin{equation}\label{W exp}
\bar{W} \sim \bar W_{-1/2}\epsilon^{-1/2} + \bar W_0 + \epsilon^{1/2} \bar W_{1/2} + \epsilon \bar W_1 + \epsilon^{3/2}\bar{W}_{3/2} + \cdots;
\end{equation}
condition \eqref{gap solid} then confirms the scaling $\lambda=O(\epsilon^{-1/2})$, and we anticipate the expansion 
\begin{equation}\label{lam expansion}
\lambda \sim \lambda_{-1/2}\epsilon^{-1/2}+\lambda_0 + \lambda_{1/2}\epsilon^{1/2}+\cdots. 
\end{equation}
The gap problem at each order is found by substitution of \eqref{W exp} into \eqref{gap lap}--\eqref{int gap} and by mapping \eqref{gap bubble} onto the nominal surface $Y=H_0(X)$ by means of a Taylor expansion in $Y$; in particular, it is readily seen from \eqref{gap lap} and \eqref{gap periodic} that $W_{-1/2}$ and $W_{0}$ are independent of $Y$, namely $W_{-1/2}=W_{-1/2}(X)$, $W_0=W_0(X)$.

In the outer region we anticipate, subject to confirmation through matching, that the disturbance velocity is $O(1)$. We accordingly expand $\bar{w}$ as
\begin{equation}\label{assumption}
\bar{w} \sim \bar{w}_0 + o(1), \quad \bar{w}_0 = O(1),
\end{equation}
where the leading-order outer problem governing $\bar{w}_0$ is shown in Fig.~\ref{fig:inner_outer}. The depicted domain is bounded by the two rays $y=\pm1$ $(x>0)$ and the semicircle $x^2+y^2=1$ $(x>0)$; the error incurred by mapping  the boundary conditions on $y=\pm (1+\epsilon)$ to $y=\pm1$ is small in $\epsilon$ and accordingly does not enter the leading-order problem. Thus, the outer disturbance velocity $\bar{w}_0$ satisfies Laplace's equation, attenuation as $x\to\infty$, periodicity at $y=\pm1$, and a boundary condition identical to \eqref{bubble bar} on the half circle. The boundary of the leading-order outer region is non-smooth where the rays and semi-circle coincide; at these points $\bar{w}_0$ is allowed to be singular, the only requirement being that matching with the gap region is satisfied.

\subsection{Leading-order asymptotics}\label{ssec:leading}
Consider the gap region. Laplace's equation \eqref{gap lap} at $O(\epsilon^{1/2})$ reads
\begin{equation}\label{W12 eq}
\frac{d^2\bar W_{-1/2}}{dX^2}+\pd{^2\bar{W}_{1/2}}{Y^2} = 0.
\end{equation}
Integrating with respect to $Y$ between $0$ and $H_0(X)$, together with the appropriate asymptotic orders of \eqref{gap periodic} and \eqref{gap bubble}, yields
\begin{equation}\label{gap W12}
\frac{d}{dX}\left(H_0\frac{d\bar W_{-1/2}}{dX}\right)= 0;
\end{equation}
this is precisely the ``flux'' conservation law anticipated in subsection A. 
Integrating, in conjunction with the conditions 
\begin{equation}
 \quad \bar{W}_{-1/2}=-\lambda_{-1/2}, \quad  \frac{d\bar{W}_{-1/2}}{dX} = 1  \quad \text{at} \quad X=0,
\end{equation}
which respectively follow from \eqref{gap solid} and \eqref{int gap}, we find 
\begin{equation}\label{leading W}
\bar W_{-1/2} = \sqrt{2}\arctan\frac{X}{\sqrt{2}}-\lambda_{-1/2}.
\end{equation}

Consider now the far-field behaviour of \eqref{leading W},
\begin{equation}\label{gap W12 far}
\bar{W}_{-1/2} \sim \frac{\pi}{\sqrt{2}} -\lambda_{-1/2} - \frac{2}{X} + O\left(\frac{1}{X^3}\right) \quad \text{as} \quad X\to \infty.
\end{equation}
According to van Dyke's matching rule \cite{Hinch:91}, the constant leading-order term in \eqref{gap W12 far} implies an $O(\epsilon^{-1/2})$ disturbance velocity in the outer region, forced solely by the condition that it approaches $\pi/\sqrt{2}-\lambda_{-1/2}$ in the limit where $(x,y)\to(0,\pm 1)$ from within the outer liquid domain. The only such solution, however,   is constant everywhere, contradicting the far-field condition that $\bar{w}$ attenuates as $x\to\infty$. 
It follows that there cannot be an $O(\epsilon^{-1/2})$ term in the outer region, thereby confirming assumption \eqref{assumption} [cf.~\eqref{matching}] and showing that 
\begin{equation}
\lambda_{-1/2} = \frac{\pi}{\sqrt{2}}.
\end{equation}

\subsection{Leading-order correction}\label{ssec:lcor}
It is readily found that the gap correction $\bar W_0$ is governed by an equation identical to \eqref{gap W12}. It follows that
\begin{equation}\label{correction W}
\bar W_{0} = C\arctan\frac{X}{\sqrt{2}}-\lambda_{0};
\end{equation}
but since the global relation \eqref{int gap} is trivial at $O(\epsilon^{1/2})$, $C=0$. The leading correction to the slip length, $\lambda_0$, is determined as follows. On one hand, given \eqref{gap W12 far} and \eqref{correction W}, van Dyke's matching rule shows that the leading-order outer field satisfies
\begin{equation}\label{matching}
\bar{w}_0 \sim - \frac{2}{x} - \lambda_{0} + o(1)  \quad \text{as} \quad (x,y)\to (0,\pm 1). 
\end{equation}
On the other hand, the outer-region problem governing $\bar{w}_0$, shown in Fig.~\ref{fig:inner_outer}, is closed by the lower-order matching condition $\bar{w}_0 \sim - {2}/{x}$. Thus once $\bar{w}_0$ is solved for, the slip-length correction can be found as $\lambda_0=-\lim_{x\to0}(\bar{w}_0+2/x)$. In the appendix we solve the outer problem using a conformal mapping, finding
\begin{equation}\label{lambda0}
\lambda_0 = -\frac{12}{\pi}\ln 2\approx -2.648.
\end{equation}

\subsection{First-order correction}
Turning again to the inner region, integration of \eqref{W12 eq} together with the $O(\epsilon^{1/2})$ balance of \eqref{gap periodic} shows that
\begin{equation}\label{W12 sol}
\bar{W}_{1/2} = \frac{1}{2}\frac{X}{H_0^2(X)}Y^2+\mathcal{A}(X),
\end{equation}
where $\mathcal{A}(X)$ is an integration constant. A solvability condition on $\bar{W}_{3/2}$ is derived in the usual way by integrating \eqref{gap lap} from $Y=0$ to $H_0(X)$ while using the $O(\epsilon^{3/2})$ balances of the periodicity condition \eqref{gap periodic} and the no-shear condition \eqref{gap bubble}, the latter balance being
\begin{equation}
\pd{\bar{W}_{3/2}}{Y} = \frac{dH_1}{dX}\frac{d\bar{W}_{-1/2}}{dX}+\frac{dH_0}{dX}\pd{\bar{W}_{1/2}}{X}-H_1\pd{^2\bar{W}_{1/2}}{Y^2}+\frac{dH_0}{dX} \quad \text{at} \quad Y=H_0(X).
\end{equation}
The resulting solvability condition provides a differential equation governing $\mathcal{A}(X)$; in conjunction with the $O(\epsilon^{3/2})$ and $O(\epsilon^{2})$ balances of \eqref{gap solid} and \eqref{int gap}, respectively, we find the following problem: 
\begin{equation}
\frac{d}{dX}\left(H_0\frac{d\mathcal{A}}{dX}\right)=\frac{2X}{(2+X^2)^2}-X, \quad \left.\mathcal{A}\right|_{X=0}=-\lambda_{1/2}, \quad \left.\frac{d\mathcal{A}}{dX}\right|_{X=0}=-1/6.
\end{equation}
From the solution to this problem it follows that 
\begin{equation}\label{A}
\mathcal{A}\sim -X + \frac{13\pi}{12\sqrt{2}}-\lambda_{1/2} + O\left(\frac{1}{X}\right) \quad \text{as} \quad X\to\infty.
\end{equation}
The leading term in \eqref{A}, along with the leading term in an expansion of the $Y$-dependent term in \eqref{W12 sol}, is expected to match with high-order terms in the inner limit of $\bar{w}_0$. 
The constant term in \eqref{A}, however, forces a constant outer-region solution at  $O(\epsilon^{1/2})$, which contradicts the attenuation of $\bar{w}$ as $x\to\infty$; note that the deviation of the periodic-cell boundaries from $y=\pm1$ modifies the outer-region problem only at $O(\epsilon)$.  We thus find
\begin{equation}
\lambda_{1/2}=\frac{13\pi}{12\sqrt{2}}\approx 2.407.
\end{equation}
\begin{figure}[t]
\begin{center}
\includegraphics[scale=0.4]{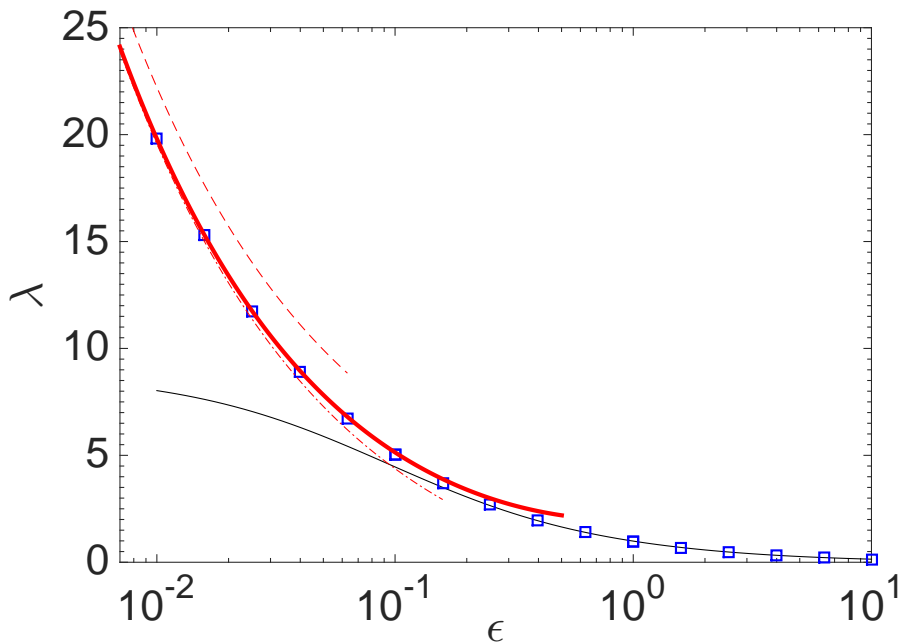}
\caption{Slip length normalised by bubble radius, as a function of half the dimensionless minimum separation between the bubbles. Thick line --- near-contact asymptotics \eqref{result}; Thin dash-dotted line --- two first terms of \eqref{result}; Thin dashed line --- leading singular term of \eqref{result}; Symbols --- numerical solution using the ``unified-transform'' method \cite{Crowdy:15b}; Thin line --- dilute-limit approximation \cite{Crowdy:16}, see Sec.~\ref{sec:disc} for details.}
\label{fig:comp}
\end{center}
\end{figure}

\section{Corroboration and discussion} \label{sec:disc}
To recapitulate, we have derived the near-contact asymptotics of the effective slip length, normalised by the bubble radius, as
\begin{equation}\label{result}
\lambda \sim \frac{\pi}{\sqrt{2\epsilon}}-\frac{12}{\pi}\ln 2 + \frac{13\pi}{12\sqrt{2}}\sqrt{\epsilon} +\cdots  \quad \text{as} \quad \epsilon\to0.
\end{equation}
Fig.~\ref{fig:comp} demonstrates excellent agreement of our asymptotic result 
with an ``exact'' numerical solution obtained using an accurate and efficient scheme derived from the ``unified-transform'' method \cite{Crowdy:15b}. Also shown is the approximation $\lambda \approx \pi l^{-1} /[2-\pi^2/(6l^2)]$, where $l=1+\epsilon$, derived in the ``dilute'' limit of well-separated bubbles \cite{Crowdy:16}. While the dilute and near-contact limit do not asymptotically overlap, they together provide a rather complete description, for arbitrary $\epsilon$, in terms of elementary expressions.  As was pointed out in \cite{Crowdy:11}, the problem considered herein can be mapped using symmetry to the potential-flow problem of calculating the ``blockage coefficient'' for a circular cylinder in a infinite slab. The present asymptotic solution may therefore have ramifications also in electrostatics, flow through porous-media, and large-Reynolds-number hydrodynamics. 

We have focused in this paper on the case where the contact angle is $\pi/2$. For contact angles appreciably below $\pi/2$, the inner region is no longer narrow, leading to a gap velocity varying over an $O(\epsilon)$ length scale rather than $O(\epsilon^{1/2})$. The divergence of the effective slip length as $\epsilon\to0$ is then logarithmic in $\epsilon$ \cite{Ybert:07}. A detailed asymptotic analysis of the effective slip length for arbitrary contact angles is underway. 


\textbf{Acknwoledgements.}  I thank Professor Darren Crowdy for pointing me to Refs.~\onlinecite{Morris:03}--\onlinecite{Ablowitz:Book} for the conformal mapping employed in the appendix, and Ms Elena Luca for sending me a Matlab code that solves for the slip length numerically based on a``unified-transform'' method \cite{Crowdy:15b}, which was used to produce the symbols in Fig.~\ref{fig:comp}. I am also grateful to Ehud Yariv for helpful suggestions and to the anonymous Referee that spotted an error in an earlier version of this paper. 

\appendix
\section{Solution to leading-order outer problem }
\begin{figure}[t]
\begin{center}
\includegraphics[scale=0.9]{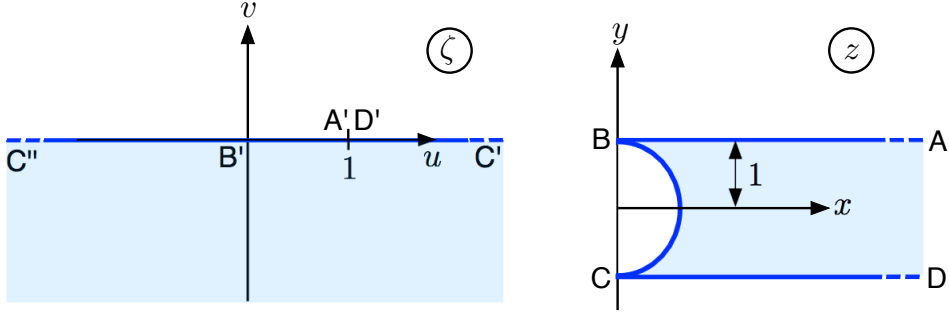}
\caption{The conformal mapping employed in the appendix.}
\label{fig:mapp}
\end{center}
\end{figure}

We here solve the leading-order outer problem as shown in Fig.~\ref{fig:inner_outer}, supplemented by the matching condition \eqref{matching} discussed in section \ref{sec:asym}.  As a preliminary step, we introduce a conformal mapping between the lower half of an auxiliary complex-plane $\zeta=u+iv$ to the zero-angle curvilinear degenerate triangle in the physical plane $z=x+iy$. Fixing the locations of the critical points on the $u$ axis as depicted in Fig.~\ref{fig:mapp}, the required mapping is written as \cite{Ablowitz:Book}
\begin{equation}\label{map}
z = i + \frac{2\Lambda(\zeta)}{\Lambda(1-\zeta)},
\end{equation}
where $\Lambda$ stands for the hypergeometric function
\begin{equation}\label{F def}
\Lambda(\zeta) = F\left(\frac{1}{2},\frac{1}{2},1,\zeta\right)=\frac{1}{\pi}\int_0^1t^{-1/2}(1-t)^{-1/2}(1-\zeta t)^{-1/2}\,dt,
\end{equation}
with $z^p=\exp[p\log(z)]$, the branch cut of the principle-value logarithm taken along the negative real axis. Note that $\Lambda(\zeta)$ is a single-valued analytical function in the $\zeta$ plane excluding the branch-cut ray $u>1$  along the real axis $v=0$ \cite{Abramowitz:book}. Along this brach cut $\Lambda(\zeta)$ is discontinuous \footnote{Relation \eqref{pm relation} can be derived following Morris \cite{Morris:03}, who split the integral in \eqref{F def} at $t=1/(1+\lambda)$, and then wrote the resulting two integrals in terms of  hypergeometric functions by changing variables. We note however that Morris's derivation ignores the discontinuity, whereby he finds only one of the limits in \eqref{pm relation}; this overlook leads him to incorrectly conclude that his mapping, which, up to a translation, is the same as \eqref{map}, should be considered from the upper (rather than lower) half plane.},
\begin{equation}\label{pm relation}
\lim_{\delta\to0}\Lambda(1+\lambda-i\delta) = \mp i\Lambda(-\lambda)+(1+\lambda)^{-1/2}\Lambda\left(\frac{1}{1+\lambda}\right) \quad \text{for} \quad \delta\gtrless0,
\end{equation}
where $\lambda>0$ is real; note that $\Lambda(u)$ is real and positive for $u<1$.

The mapping \eqref{map} is verified as follows. First, note that $\Lambda(u)$, where $0<u<1$, is real and positive, ranging from $1$ to $\infty$; it follows that $\Lambda(u)/\Lambda(1-u)$ spans the positive real axis and hence from \eqref{map} that $A'B'$ is mapped to $AB$ (see Fig.~\ref{fig:mapp}). Next, using \eqref{pm relation} and \eqref{map} we find
\begin{gather}\label{limit1}
\lim_{\delta\to0}z(1+\lambda-i\delta)= -i+\frac{2\Lambda\left(\frac{1}{1+\lambda}\right)}{(1+\lambda)^{1/2}\Lambda(-\lambda)}, \\ \label{limit2}
\lim_{\delta\to0}z(-\lambda-i\delta)= i+\frac{2\Lambda(-\lambda)}{i\Lambda(-\lambda)+(1+\lambda)^{-1/2}\Lambda\left(\frac{1}{1+\lambda}\right)},
\end{gather}
where $\lambda$ and $\delta$ are positive and real. In \eqref{limit1}, the second term on the right hand side spans the positive real axis, and therefore $C'D'$ (approached from the lower half plane) is mapped to $CD$. Finally, it is readily verified that the absolute magnitude of \eqref{limit2} is unity, showing that $B'C''$ is mapped to the semi-circle $BC$. 

We now look for a solution in the form $\bar{w}_0+x = \mathrm{Re}\{T(\zeta)\}$, where $T$ is an analytical function in the half-plane $\mathrm{Im}\{\zeta\}<0$. To this end, we invoke the asymptotic relations \cite{Abramowitz:book}
\begin{equation}\label{AS}
\Lambda(\zeta)\sim -\frac{1}{\pi}\log(1-\zeta)+\frac{1}{\pi}\ln 16 + o(1) \quad \text{as} \quad \zeta\to1,  \quad \Lambda(\zeta)\sim 1+o(1) \quad \text{as} \quad \zeta\to0,
\end{equation}
where in the first $|\arg(1-\zeta)|<\pi$; the corresponding behaviours of $\Lambda(1-\zeta)$ as $\zeta\to0,1$ readily follow. Together with \eqref{map}, the above asymptotic relations imply that the far-field condition $\lim_{x\to\infty}\bar{w}_0=0$ and the matching condition, that  $\bar{w}_0\sim -2/x$ as $x\to0$, with $(1-y)\ll x$, are satisfied if
\begin{equation}\label{mapping conditions} 
T\sim z + o(1) \quad \text{as} \quad \zeta\to1, \quad  T\sim -\frac{2}{z-i}+O(1) \quad \text{as} \quad \zeta\to0.
\end{equation}
Employing \eqref{map} and \eqref{AS}, it is readily verified that an analytic function satisfying \eqref{mapping conditions} for which $\mathrm{Re}\{T\}$ satisfies Neumann conditions on the domain boundary is
\begin{equation}
T=\frac{1}{\pi}\log\zeta -\frac{2}{\pi}\log(1-\zeta)+i+\frac{8}{\pi}\ln 2.
\end{equation}
Inspecting the limit as $\zeta\to0$, and using \eqref{AS}, we find
\begin{equation}
\bar{w}_0 \sim -\frac{2}{x} + \frac{12}{\pi}\ln 2 + o(1) \quad \text{as} \quad x\to0 \quad (y-1\ll x),
\end{equation}
from which the result \eqref{lambda0} follows.

\bibliography{refs.bib}

\end{document}